\documentclass[prd,a4paper,preprint,superscriptaddress,nofootinbib]{revtex4-2}

\usepackage[hyperfootnotes=false]{hyperref}
\hypersetup{linkcolor=red,
citecolor=green,
filecolor=cyan,
urlcolor=magenta}

\usepackage{amsfonts}

\usepackage{amsmath}
\usepackage{amssymb}

\usepackage{bbm}

\usepackage{bbold}

\usepackage{adjustbox}
\usepackage{booktabs}

\usepackage{colortbl}

\usepackage{cancel}

\usepackage{color}
\usepackage{xcolor}

\usepackage{graphicx}
\usepackage{xspace}

\usepackage{units}

\usepackage{slashed}

\usepackage{tensor}

\usepackage{gensymb}

\usepackage{cleveref}

\usepackage{tabularx}
\usepackage{multirow}

\usepackage{setspace}
\usepackage{orcidlink}

\usepackage{caption}
\usepackage{subcaption}

\usepackage{enumerate}
\usepackage{enumitem}

\usepackage[version=3]{mhchem}

\usepackage{lipsum}

\newcommand{{\ia} }{{\i}}
\newcommand{{\Ia} }{{\.I}}

\def\s2tw{{\rm sin ^2 \theta_{W}}}

\def\ds{\displaystyle}
\def\beq{\begin{equation}}
\def\eeq{\end{equation}}
\def\bea{\begin{eqnarray}}
\def\eea{\end{eqnarray}}
\def\ve{\vert}
\def\vel{\left|}
\def\ver{\right|}
\def\nnb{\nonumber}
\def\lla{\left<}
\def\rra{\right>}
\def\nnb{\nonumber}

\def\es{ &=& }
\def\ar{&+& }
\def\ek{&-& }
\def\cp{&\times&}

\def\ds{\displaystyle}
\def\bea{\begin{eqnarray}}
\def\eea{\end{eqnarray}}
\def\beeq{\begin{eqnarray}}
\def\eeeq{\end{eqnarray}}
\def\ve{\vert}
\def\vel{\left|}
\def\ver{\right|}
\def\nnb{\nonumber}

\def\lla{\left<}
\def\rra{\right>}

\def\nnb{\nonumber}

\def\ba{\begin{array}}
\def\ea{\end{array}}

\def\xis0{{\Xi^{*0}}}

\def\g5{\gamma_5}

\def\olra{\stackrel{\leftrightarrow}}
\def\ola{\stackrel{\leftarrow}}
\def\ora{\stackrel{\rightarrow}}

\begin{document}


\title{Determination of the  multipole moments of the  $J^P = 3^-$ mesons via QCD sum rules}

\author{T.~M.~Aliev\,\orcidlink{0000-0001-8400-7370}}
\email{taliev@metu.edu.tr}
\affiliation{Department of Physics, Middle East Technical University, Ankara, 06800, Turkey}

\author{S.~Bilmis\,\orcidlink{0000-0002-0830-8873}}
\email{sbilmis@metu.edu.tr}
\affiliation{Department of Physics, Middle East Technical University, Ankara, 06800, Turkey}
\affiliation{TUBITAK ULAKBIM, Ankara, 06510, Turkey}

\author{M.~Savci\,\orcidlink{0000-0002-6221-4595}}
\email{savci@metu.edu.tr}
\affiliation{Department of Physics, Middle East Technical University, Ankara, 06800, Turkey}

\date{\today}

\begin{abstract}

  The multipole moments of the  $J^P = 3^-$ tensor mesons are obtained using light-cone sum rules. Our results are compared with the predictions of the light-cone helicity formalism within the framework of SU(3) flavor symmetry. We observe a good agreement between the two approaches. The obtained multipole moments of tensor mesons provide deeper insights into their properties and better understanding of the strong and electromagnetic decay channels of these mesons.
  
\end{abstract}
\maketitle
%
%
%
\newpage
%

\section{Introduction}

Detailed studies of the properties of well-known light mesons, as well as searches for new meson states, constitute a primary focus of many experiments. Light meson states with $J^P = 3^-$ have been observed in experiments conducted by various collaborations \cite{Baldi:1976ua,Brandenburg:1975ft,Wagner:1974gw,Aston:1988rf}. Heavy tensor mesons are investigated in ongoing experiments such as LHCb \cite{LHCb:2018roe}, Compass
\cite{Ketzer:2019wmd}, Gluex \cite{Austregesilo:2018mno},
BES III \cite{BESIII:2020nme,Mezzadri:2015lrw,Marcello:2016gcn},
Panda \cite{Fioravanti:2012px}. A comprehensive analysis of the current status of the tensor mesons is given in \cite{Ketzer:2019wmd,Padmanath:2018zqw}.

After the discovery of the tensor mesons, the next step is to study their properties in detail. The study of electromagnetic form factors and multipole moments plays a critical role in understanding the inner structure of these states. Electromagnetic form factors of mesons are widely studied in the literature. For example, the electromagnetic form factors of the $\pi$ meson have been studied extensively (see \cite{Baldi:1976ua,Brandenburg:1975ft,Wagner:1974gw,Aston:1988rf,Ketzer:2019wmd,LHCb:2018roe} and the references therein). The form factors of the vector mesons are studied in \cite{BESIII:2020nme,Mezzadri:2015lrw,Marcello:2016gcn,Austregesilo:2018mno,Fioravanti:2012px}, in lattice QCD \cite{LHCb:2018roe,BESIII:2020nme,Padmanath:2018zqw}. The magnetic and quadrupole moments of heavy vector and axial vector mesons are studied using light cone sum rules in \cite{Aliev:2023tqy} (for details on light cone sum rules, see \cite{Balitsky:1987bk}). The magnetic dipole moment of the light tensor mesons is studied in \cite{Aliev:2009np} within the framework of light cone sum rules.

In the present work, we study the multipole moments of the light $J^P = 3^-$ tensor mesons. The paper is organized as follows. In section~\ref{sec:2}, the light cone sum rules for the multipole moments of the light $J^P = 3^-$ tensor mesons are derived. Section~\ref{sec:3} presents a numerical analysis of the sum rules for the multipole moments, and our conclusions.
\section{Light cone sum rules for the multipole moments of $J^P = 3^-$
tensor mesons}
\label{sec:2}
In this section, we study the multipole moments of the $J^P = 3^-$ tensor mesons. The transition matrix element of the electromagnetic current between $J^P = 3^-$ tensor states is defined in terms of the form factors as follows~\cite{Lorce:2009br}:

\bea
\label{eq1}
\lla 3^-
(p^\prime) \vel j^\mu \ver 3^- (p) \rra \es 
\varepsilon_{\alpha_1^\prime \alpha_2^\prime \alpha_3^\prime}^\ast (p^\prime)
\Bigg\{ {\cal P}^\mu\Bigg[g^{\alpha_1^\prime \alpha_1} g^{\alpha_2^\prime
\alpha_2} g^{\alpha_3^\prime \alpha_3} F_1(Q^2) \nnb \\
\ar  {q^{\alpha_1^\prime}q^{\alpha_1} \over 2 m_3^2} 
{q^{\alpha_2^\prime}q^{\alpha_2} \over 2 m_3^2}
g^{\alpha_3^\prime \alpha_3}  F_5(Q^2) -
 {q^{\alpha_1^\prime}q^{\alpha_1} \over 2 m_3^2} 
g^{\alpha_2^\prime \alpha_2}g^{\alpha_3^\prime \alpha_3}  F_3(Q^2) \nnb \\
\ek
 {q^{\alpha_1^\prime}q^{\alpha_1} \over 2 m_3^2} 
 {q^{\alpha_2^\prime}q^{\alpha_2} \over 2 m_3^2} 
 {q^{\alpha_3^\prime}q^{\alpha_3} \over 2 m_3^2}  F_7(Q^2) \Bigg] \nnb \\
\ar (g^{\mu\alpha_3}  q^{\alpha_3^\prime} - 
g^{\mu\alpha_3^\prime} q^{\alpha_3}) \Bigg[ 
g^{\alpha_1^\prime \alpha_1} g^{\alpha_2^\prime \alpha_2}  F_2(Q^2) -
 {q^{\alpha_1^\prime}q^{\alpha_1} \over 2 m_3^2} 
g^{\alpha_2^\prime \alpha_2}  F_4(Q^2) \nnb \\
\ar  {q^{\alpha_1^\prime}q^{\alpha_1} \over 2 m_3^2} 
 {q^{\alpha_2^\prime}q^{\alpha_2} \over 2 m_3^2}  F_6(Q^2)
\Bigg] \Bigg\} \varepsilon_{\alpha_1 \alpha_2 \alpha_3} (p)
\eea
where $p(p^\prime)$ and $\varepsilon_{\alpha \beta \gamma}$ are the
momentum and polarization tensor of the tensor meson, ${\cal P}^\mu =
(p+p^\prime)^\mu$, $F_i$ are the form factors, and $m_3$ is the mass of the corresponding tensor meson.

From an experimental point of view, the multipole form factors are more convenient to measure. The relations between the two sets of form factors for the arbitrary integer spin (as well as arbitrary half spin) case can be found in \cite{Lorce:2009br}. For the real photon case, i.e., $q^2=0$, these relations are given as:
\bea
\label{eq2}
F_{2 k +1}(0) \es \sum_{\ell = 0}^k C_{j-\ell}^{j-k} (-1)^{k-\ell} 
\Big[G_{E_{2 \ell}} (0) + (1-\delta_{\ell 0}) [G_{M_{2 \ell-1 }} (0)] \Big]~, \nnb \\
F_{2 k +2}(0) \es \sum_{\ell = 0}^k C_{j-\ell-1}^{j-k-1} (-1)^{k-\ell}
G_{M_{2 \ell+1}} (0) ~,
\eea
where
\begin{equation*}
\label{eq2}
    \begin{aligned}
     C_n^k =
        \begin{cases}
{\ds n!\over \ds  k! (n-k)!} & {\rm when~} n \ge k \ge 0~, \\
      0 & {\rm otherwise}~. \\
        \end{cases}
    \end{aligned}
\end{equation*}
To derive the light cone sum rules for the multipole moments, we start by considering the following correlation function:
\bea
\label{eq3}
\Pi_{\alpha\beta\rho\lambda\sigma\tau}^\mu = - \int d^4x \int d^4y e^{ipx + iqy}
\lla 0 \vel {\rm T} \Big\{ j_{\alpha\beta\rho}(x) j_{el}^\mu (y)
j_{\lambda\sigma\tau} (0) \Big\} \ver 0 \rra ~,
\eea
where $j_{el}^\mu$ is the electromagnetic current for the light quarks
given as:
\bea
j_{el}^\mu = e_u \bar{u} \gamma^\mu u +  e_d \bar{d} \gamma^\mu d + 
e_s \bar{s} \gamma^\mu s~,\nnb
\eea
$e_q$ is the electric charge of the corresponding quarks, and
\bea
\label{eq4}
j_{\alpha_1\alpha_2\alpha_3} = {1\over 6} \bar{q}
\Gamma_{\alpha_1\alpha_2\alpha_3} q~,
\eea
is the interpolating current of the $J^P = 3^-$
tensor mesons (For the form of the interpolating current see also~\cite{Wang:2016hoi}). Here
\bea
\label{eq5}
\Gamma_{\alpha\beta\rho} \es
\gamma_\alpha \Big(  \olra{\cal D}_{\beta} \olra{\cal D}_{\rho} +
\olra{\cal D}_{\rho} \olra{\cal D}_{\beta} \Big) +
\gamma_\beta \Big(  \olra{\cal D}_{\alpha} \olra{\cal D}_{\rho} +
\olra{\cal D}_{\rho} \olra{\cal D}_{\alpha} \Big) +
\gamma_\rho \Big(  \olra{\cal D}_{\alpha} \olra{\cal D}_{\beta} +
\olra{\cal D}_{\beta} \olra{\cal D}_{\alpha} \Big)~,
\eea
where
\bea
\label{eq6}
\olra{\cal D}_{\alpha} \es {1\over 2} \left( \ora{\cal D}_{\alpha} - \ola{\cal D}_{\alpha}\right)~, \nnb \\
\ora{\cal D}_{\alpha} \es \ora{\partial}_\alpha - {i\over 2} g A_\alpha^a \lambda^a~,\nnb \\
\ola{\cal D}_{\alpha} \es \ola{\partial}_\alpha + {i\over 2} g A_\alpha^a \lambda^a~,
\eea
and $\lambda^a$ are the Gell Mann  matrices. Introducing an electromagnetic
background field of a plane wave $F_{\mu\nu} = i (\varepsilon_\mu
q_\nu - \varepsilon_\nu q_\mu) e^{iqx}$, the correlator function can be written as:
\bea
\label{eq7}
\varepsilon^\mu\Pi_{\alpha\beta\rho \mu \lambda\sigma\tau} = i \int d^4x e^{ipx}
\lla 0 \vel {\rm T} \Big\{ j_{\alpha\beta\rho}(x)
j_{\lambda\sigma\tau} (0) \Big\} \ver 0 \rra_F ~,
\eea
where subscript $F$ means that the vacuum expectation value is evaluated in
the background field. It should be noted here here that by expanding Eq.
(\ref{eq7}) in powers of $F_{\mu\nu}$ and retaining only the linear term in
$F_{\mu\nu}$, the correlation function given in Eq. (\ref{eq3}) can easily
be reproduced.

According to QCD sum rules methodology, the correlation function given in Eq.
(\ref{eq3}) should be calculated in two different kinematical domains. In
one domain, Eq. (\ref{eq7}) is dominated by the $T \to T \gamma$ transition
if the incoming and outgoing states are close to the tensor meson shell, i.e.,
$p^2\simeq m_T^2$, and $p^{\prime 2}\simeq m_T^2$. On the other hand, the
correlation functions calculated in deep Euclidean region where $p^2\ll 0$,
$p^{\prime 2} \ll 0$ using the operator product expansion (OPE) in terms of
the photon distribution amplitudes (DAs) with increasing twist.
At the hadronic level, Eq. (\ref{eq3}) can be written as:
\bea
\label{eq8}
\varepsilon^\mu \Pi_{\alpha\beta\rho\mu\lambda\sigma\tau} =
\sum_{i=0,1,2,3}
{\varepsilon^\mu \lla 0 \vel j_{\alpha\beta\rho}\ver
i^-(p^\prime) \rra \lla i^-(p^\prime) \vel j_\mu \ver j^-(p) \rra \lla j^-(p) \vel
j_{\lambda\sigma\tau} \ver 0 \rra \over \left(m_T^2 - p^{\prime
2}\right)\left(m_T^2-p^2\right)} + \cdots
\eea
where dots describe the contributions from excited states and continuum.
The matrix element $\lla 3^-(p^\prime) \vel j^\mu \ver 3^-(p) \rra$ is given
in Eq. (\ref{eq1}) and the remaining matrix element $\lla 0 \vel
j_{\alpha\beta\rho} \ver 3^-(p) \rra$ is defined as:
\bea
\label{eq9}
\lla 0 \vel j_{\alpha\beta\rho} \ver 3^-(p) \rra = f_3 m_3^4~
\varepsilon_{\alpha\beta\rho}^{(\lambda)}(p)~,
\eea
where $\varepsilon_{\alpha\beta\rho}^{(\lambda)}(p)$ is the polarization tensor of rank-3 tensor field, $m_3$ is its mass and $f_3$ is its decay constant.  This tensor satisfies the  following conditions:
  \begin{itemize}
  \item It is symmetric under the exchange of  any pair of indices i.e., $\varepsilon_{\alpha\beta\rho}^{(\lambda)}(p) = \varepsilon_{\beta \alpha \rho}^{(\lambda)}(p)$
  \item $\varepsilon_{\alpha\rho}^{\alpha (\lambda)}(p) = 0$
    \item It satisfies the transversality condition, $p^\alpha \varepsilon_{\alpha\beta\rho}^{(\lambda)}(p) = 0$
  \end{itemize}

It should be noted here that the current $j_{\alpha\beta\rho}$
interacts not only with $J^P = 3^-$ state but also with the states 
$J^P = 2^+$, $1^-$ and $0^+$. These matrix elements are defined as
\bea
\label{eq10}
\lla 0 \vel j_{\alpha\beta\rho} \ver 2^+(p) \rra \es f_2 m_2^2 
\left[ p_{\alpha} \varepsilon_{\beta \rho}^{(\lambda)} +
       p_{\beta} \varepsilon_{\alpha \rho}^{(\lambda)} +
       p_{\rho} \varepsilon_{\alpha \beta}^{(\lambda)} \right]~,\nnb \\
\lla 0 \vel j_{\alpha\beta\rho} \ver 1^-(p) \rra \es f_1 m_1
\left[ p_{\alpha} p_{\beta} \varepsilon_{\rho}^{(\lambda)} +
       p_{\beta} p_{\rho} \varepsilon_{\alpha}^{(\lambda)} +
       p_{\alpha} p_{\rho} \varepsilon_{\beta}^{(\lambda)} \right]~,\nnb \\
\lla 0 \vel j_{\alpha\beta\rho} \ver 0^+(p) \rra \es 
f_0 p_{\alpha} p_{\beta} p_{\rho}~,
\eea
where $f_i$ are the decay constants, and
$\varepsilon_{\alpha\beta\rho}^{(\lambda)}$,
$\varepsilon_{\alpha\beta}^{(\lambda)}$,
$\varepsilon_{\alpha}^{(\lambda)}$ are the polarization tensors of the
corresponding mesons.

To derive the completeness relation for the polarization tensor,one must first construct the general form that includes all possible combinations of the metric tensor and projection operators: $T_{\mu \nu} = g_{\mu \nu} - \frac{k_\mu k_\nu}{k^2}$ and $P_{\mu \nu} = \frac{k_\mu k_\nu}{k^2}$.

By employing symmetry arguments as well as the orthonormalization condition for the polarization tensors, $\varepsilon_{\alpha\beta\rho}^{(\lambda)}(p) \varepsilon^{\alpha\beta\rho(\lambda^\prime)}(p) = -\delta_{\lambda \lambda^\prime}$, the completeness relation for the polarization tensor can be derived. A detailed derivation of the completeness relation  is provided in~\cite{Jafarzade:2021vhh}. (See also~\cite{Wang:2016hoi,Zhu:1999pu}). The final expression of the completeness relation for spin-3 particles is given as follows:

\bea
\label{eq11}
{\cal T}_{\mu\nu\rho\alpha\beta\sigma} \es \sum_{\lambda=-3}^3
\varepsilon_{\mu\nu\rho}^{(\lambda)} \varepsilon_{\alpha\beta\sigma}^{(\lambda)} \nnb \\
\es  {1\over 15}
\Bigg[ T_{\mu \nu} (T_{\beta \sigma} T_{\rho \alpha} + T_{\alpha \sigma} T_{\rho \beta} +
      T_{\alpha \beta} T_{\rho \sigma}) +
      T_{\mu \rho} (T_{\beta \sigma} T_{\nu \alpha} + T_{\alpha\sigma} T_{\nu \beta} +
      T_{\alpha \beta} T_{\nu \sigma}) \nnb \\
\ar
      T_{\nu \rho} (T_{\beta \sigma} T_{\mu \alpha} + T_{\alpha \sigma} T_{\mu \beta} +
      T_{\alpha \beta} T_{\mu \sigma})\Bigg] \nnb \\
\ek {1\over 6}
\Bigg[ T_{\mu \alpha} (T_{\nu \sigma} T_{\rho \beta}  + T_{\nu \beta}  T_{\rho \sigma}) +
      T_{\mu \beta} (T_{\nu \sigma}  T_{\rho \alpha} + T_{\nu \alpha} T_{\rho \sigma}) +
      T_{\mu \sigma} (T_{\nu \beta}  T_{\rho \alpha} + T_{\nu \alpha} T_{\rho \beta}) \Bigg]~, \nnb \\
{\cal T}_{\mu\nu\alpha\beta} \es \sum_{\lambda=-2}^2
\varepsilon_{\mu\nu}^{(\lambda)} \varepsilon_{\alpha\beta}^{(\lambda)}
= {1\over 2} \Bigg[T_{\mu \alpha} T_{\nu \beta}  + T_{\mu \beta} T_{\nu \alpha}
- {1\over 3} T_{\mu \nu} T_{\alpha \beta}\Bigg]~, \nnb \\
{\cal T}_{\mu\nu} \es \sum_{\lambda=-1}^1
\varepsilon_\mu^{(\lambda)} \varepsilon_\nu^{(\lambda)} = T_{\mu \nu}~,
\eea
The physical part of the correlation function can be directly obtained by applying Eqs.(\ref{eq1}), (\ref{eq3}) and (\ref{eq8}-\ref{eq11}). After performing the necessary calculations, we get
\bea
\label{eq12}
\varepsilon^\mu \Pi_{\alpha\beta\rho\mu\lambda\sigma\tau} \es
{f_3^2 m_3^8 \over m_3^2-p^{\prime 2}} \varepsilon^\mu 
{\cal T}_{\alpha\beta\rho\alpha_1^\prime\alpha_2^\prime\alpha_3^\prime} (p^\prime)
{1\over m_3^2-p^2} (-1) \Bigg\{ {\cal P}^\mu \Bigg[
g^{\alpha_1\alpha_1^\prime} g^{\alpha_2\alpha_2^\prime} 
g^{\alpha_3\alpha_3^\prime}G_{E_0} \nnb \\
\ar \Bigg(- {q^{\alpha_1^\prime}q^{\alpha_1} \over 2 m_3^2} \Bigg)
g^{\alpha_2\alpha_2^\prime} g^{\alpha_3\alpha_3^\prime}
\left(-3 G_{E_0} + G_{E_2} + G_{M_1} \right) \nnb \\
\ar \Bigg(- {q^{\alpha_1^\prime}q^{\alpha_1} \over 2 m_3^2} \Bigg)
    \Bigg(- {q^{\alpha_2^\prime}q^{\alpha_2} \over 2 m_3^2} \Bigg)
g^{\alpha_3\alpha_3^\prime}
\left[3 G_{E_0} - 2 \left(G_{E_2} + G_{M_1}\right) + G_{E_4} +
G_{M_3}\right] \nnb \\
\ar \Bigg(- {q^{\alpha_1^\prime}q^{\alpha_1} \over 2 m_3^2} \Bigg)
    \Bigg(- {q^{\alpha_2^\prime}q^{\alpha_2} \over 2 m_3^2} \Bigg)
    \Bigg(- {q^{\alpha_3^\prime}q^{\alpha_3} \over 2 m_3^2} \Bigg) \nnb \\
\cp \left(- G_{E_0} + G_{E_2} + G_{M_1} - G_{E_4} -
G_{M_3} + G_{E_6} + G_{M_5} \right) \Bigg] \nnb \\
\ar \left( g^{\mu\alpha_3^\prime} q^{\alpha_3}
- g^{\mu\alpha_3} q^{\alpha_3^\prime} \right) \Bigg[
g^{\alpha_1\alpha_1^\prime} g^{\alpha_2\alpha_2^\prime} G_{M_1} +
\Bigg(- {q^{\alpha_1^\prime}q^{\alpha_1} \over 2 m_3^2} \Bigg)  
g^{\alpha_2\alpha_2^\prime} \left(G_{M_3} - 2 G_{M_1} \right) \nnb \\
\ar    \Bigg(- {q^{\alpha_1^\prime}q^{\alpha_1} \over 2 m_3^2} \Bigg)
    \Bigg(- {q^{\alpha_2^\prime}q^{\alpha_2} \over 2 m_3^2} \Bigg)
 \left( G_{M_1} - G_{M_3} + G_{M_5} \right) \Bigg] \Bigg\}
{\cal T}_{\alpha_1\alpha_2\alpha_3\lambda\sigma\tau} (p)~,
\eea
where ${\cal P}_\mu = (p+p^\prime)_\mu$ and $\epsilon^\mu$ is the polarization vector of a photon field. 
From Eq. (\ref{eq10}) and (\ref{eq11}), we see that the terms containing
maximum number of $g_{\alpha\mu}$ contain contributions only from 
$J^P = 3^-$ states. For this reason we retain only these terms in 
$\varepsilon^\mu {\cal T}_{\mu\nu\rho\alpha\beta\sigma}$ 
and contributions coming from other terms that involve
$J^P = 2^+$, $1^-$, $0^+$ states are not considered.
From Eq. (\ref{eq12}), it is evident that the correlation function contains many
structures. To determine the multipole moments $G_{M_1}$, $G_{M_3}$,
$G_{M_5}$, $G_{M_7}$
as well as $G_{E_0}$, $G_{E_2}$, $G_{E_4}$, $G_{E_6}$, we choose the Lorentz structures listed in Table~\ref{tab:1}.
%
\begin{table}[h]

\renewcommand{\arraystretch}{1.3}
\addtolength{\arraycolsep}{-0.5pt}
\small
$$
\begin{array}{c|l}
    \toprule
            \mbox{Invariant} & {\multirow{2}{*}{\mbox{Structures}}}  \\
            \mbox{functions}   & \\
    \midrule
\Pi_1 & ({\cal P}\!\cdot\!\varepsilon) g_{\alpha\rho} 
g_{\beta\nu} g_{\mu\sigma} \\
\Pi_{2} & ({\cal P}\!\cdot\!\varepsilon) q_{\rho} q_{\sigma} 
g_{\alpha\nu} g_{\beta\mu}  \\
\Pi_{3} & ({\cal P}\!\cdot\!\varepsilon) q_{\beta} q_{\nu}
q_{\rho} q_{\sigma} g_{\alpha\mu}  \\
\Pi_{4} & ({\cal P}\!\cdot\!\varepsilon) q_{\alpha} q_{\beta}
q_{\mu} q_{\nu} q_{\rho} q_{\sigma} \\
\Pi_{5} & \varepsilon_{\sigma} q_{\rho} g_{\alpha\nu}
g_{\beta\mu} \\
\Pi_{6} & \varepsilon_{\sigma} q_{\rho} g_{\alpha\beta}
g_{\mu\nu} \nnb\\
\Pi_{7} & \varepsilon_{\rho} q_{\sigma} g_{\alpha\beta}
g_{\mu\nu} \nnb\\
\Pi_{8} & \varepsilon_{\sigma} q_{\beta} q_{\nu}
q_{\rho} g_{\alpha\mu} \\
\Pi_{9} & \varepsilon_{\rho} q_{\beta} q_{\nu}
q_{\sigma} g_{\alpha\mu} \\
\Pi_{10} & \varepsilon_{\rho} q_{\alpha} q_{\beta}
q_{\sigma} g_{\mu\nu} \\
\Pi_{11} & \varepsilon_{\sigma} q_{\alpha} q_{\beta}
q_{\mu} q_{\nu} q_{\rho} \\
    \bottomrule
\end{array}
$$
\caption{Invariant functions $\Pi_i$ and their corresponding structures.}
\renewcommand{\arraystretch}{1}
\addtolength{\arraycolsep}{-1.0pt}
\label{tab:1}
\end{table}
%

As noted earlier, the QCD side of the correlation function can be
calculated in the deep Euclidean domain where $p^{\prime 2} \to -\infty$ and
$p^2 \to\infty$ using the operator product expansion. To calculate
the correlation function in the deep Euclidean space, it is necessary to insert
the expressions of the interpolating current $j_{\alpha\beta\rho}$ into Eq.
(\ref{eq7}), resulting in:
\bea
\label{eq14}
\varepsilon^\mu \Pi_{\alpha\beta\rho\mu\lambda\sigma\tau} \es {i\over 36} \int d^4x
e^{i p^\prime (x-y)} \lla 0 \vel {\rm T} \big\{ \bar{q}_1(x)
\Gamma_{\alpha\beta\rho}q_2(x) \bar{q}_2(y)
\Gamma_{\lambda\sigma\tau} q_1(y) \big\} \ver 0\rra \ve_{{y=0},F}
\eea

Applying the Wick theorem, the light quark operators appear in the presence
of the gluonic and electromagnetic background fields. The
expression of the light quark propagator in the presence of the background
field is given as \cite{Balitsky:1987bk}:
\bea
\label{eq15}
iS_q(x) \es
{i \not\!x \over 2\pi^2 x^4} - {i g \over 16 \pi^2 x^2} \int_0^1 du
\Big\{\bar{u} \rlap/x \sigma^{\alpha\beta} + u \sigma^{\alpha\beta} \rlap/x
\Big\} G^{\alpha\beta} (ux) \nnb \\
\ek {i e_q \over 16 \pi^2 x^2} \int_0^1 du
\Big\{\bar{u} \rlap/x \sigma^{\alpha\beta} + u \sigma^{\alpha\beta} \rlap/x
\Big\}{\cal F}^{\alpha\beta} (ux)~,
\eea
where $\bar{u} = 1-u$, $G^{\alpha\beta}$ and ${\cal F}^{\alpha\beta}$ are
the gluon and electromagnetic field strength tensors, respectively. To include long-distance contributions, i.e., photon radiation at long distances, it is enough to replace one of the quark propagators with:
\bea
S_{\alpha \beta}^{ab} (x-y) = - {1\over 4} \left(\Gamma_j\right)_{\alpha\beta}
\bar{q}^a (x) \Gamma_j q^b (y)~,\nnb
\eea
where 
\bea
\Gamma_j = \Big\{ I, \gamma^\mu, \gamma_5, \gamma_5 \gamma_\mu,
\sigma_{\mu\nu}/\sqrt{2} \Big\}~,\nnb
\eea
are the full set of Dirac matrices. As a result of this replacement, a photon
interacts with quark fields at long distances, and there appear matrix
elements of non local operators such as $\bar{q}(x) \Gamma q^\prime (0)$
and $\bar{q}(x) G_{\mu\nu} q^\prime (0)$ between vacuum states in the
presence of the background fields. These matrix elements are defined in
terms of photon DAs and given in \cite{Ball:2002ps}.

The sum rules for the multipole moments can be obtained by equating the
hadronic representation of the correlation function given in Eq.
(\ref{eq12}) to the QCD side of the correlation function. Separating the
coefficients of the structures given in Table \ref{tab:1},
we get the desired sum rules for the multipole
moments in both representations of
the correlation function as given below:
\bea
\label{eq16}
G_{E_0} \es - {6 e^{m_3^2/M^2} \over f_3^2 m_3^8} \Pi_{1} \nnb \\
G_{E_2} \es - {18 e^{m_3^2/M^2} \over f_3^2 m_3^8} \Big[ \Pi_{1} - 
                2 m_3^2 \Pi_{2} \Big]
            - {225 e^{m_3^2/M^2} \over 4 f_3^2 m_3^8} \Pi_{6} \nnb \\
G_{E_4} \es - { 18 e^{m_3^2/M^2} \over f_3^2 m_3^8} \Big[
               \Pi_{1} - 4 m_3^2 \Pi_{2} + 2 m_3^4 \Pi_{3} 
             + 2 m_3^2 \Pi_{8} \Big] -
            {225 e^{m_3^2/M^2} \over 2 f_3^2 m_3^8} \Pi_{6} \nnb \\
G_{E_6} \es - {2 e^{m_3^2/M^2} \over f_3^2 m_3^8} \Big[
    3 \Pi_{1} - 18 m_3^2 \Pi_{2} + 18 m_3^4 \Pi_{3} - 4 m_3^6 \Pi_{4} -
    9 \Pi_{5}  - 18 m_3^2 \Pi_{9}
    - 6 m_3^4 \Pi_{11} \Big] \nnb \\
G_{M_1} \es   - {225 e^{m_3^2/M^2} \over 4 f_3^2 m_3^8} \Pi_{7} \nnb \\
G_{M_3} \es   {225 e^{m_3^2/M^2} \over 2 f_3^2 m_3^8} \Pi_{6} - 
              {36 e^{m_3^2/M^2} \over f_3^2 m_3^6} \Pi_{9}\nnb \\
G_{M_5} \es - {225 e^{m_3^2/M^2} \over 4 f_3^2 m_3^8} \Pi_{7} +
     {30 e^{m_3^2/M^2} \over f_3^2 m_3^6} \Pi_{10} 
 - {12 e^{m_3^2/M^2} \over f_3^2 m_3^4} \Pi_{11} \nnb 
\eea
where $\Pi_{1}$, $\Pi_{2}$, $\Pi_{3}$, $\Pi_{4}$,
$\Pi_{5}$, $\Pi_{6}$, $\Pi_{7}$,
$\Pi_{8}$, $\Pi_{9}$, $\Pi_{10}$, and $\Pi_{11}$ are given in the Appendix A.

The contributions of the excited states and continuum are taken into account using the quark-hadron duality
ansatz. The final steps of the sum rules involve performing the Borel transformation which suppresses these contributions.

It is important to note that the Borel parameters $M_1^2$ and $M_2^2$
are taken to be equal to each other, since initial and final states are the
same. Therefore, we set $M_1^2 = M_2^2 = 2 M^2$, and $u_0 = 1/2$, corresponding
to the case when quark and antiquark carry half of the photon momentum.

\section{Numerical analysis}
\label{sec:3}
This section presents our numerical analysis of the multipole moments. The input parameters used in calculations  are summarized in Table~\ref{tab:inputs}. For the mass and decay constants of the light tensor mesons, we adopt the values obtained from sum rule calculations as presented in~\cite{Aliev:2023tqy}. The key nonperturbative parameters for the light cone sum rules are the photon distribution amplitudes, whose explicit forms are given in \cite{Ball:2002ps}.
\begin{table}[htbp]
  \centering
  \renewcommand{\arraystretch}{1.3} 
  \setlength{\tabcolsep}{6pt} 
    \caption{Input Parameters Used in the Sum Rules}
    \begin{tabular}{@{}ll@{}}
        \toprule
        Parameter & Value \\ 
        \midrule
        $\langle \bar{q}q \rangle (1~\text{GeV})$ & $-(0.246^{+0.028}_{-0.019})^3~\text{GeV}^3$~\cite{Khodjamirian:2009ys} \\
      $\langle \bar{s}s \rangle (1~\text{GeV})$ & $(0.8 \pm 0.1 ) \langle \bar{q}q \rangle (1~\text{GeV})$~\cite{Ioffe:2002ee} \\
      $m_0^2$  & $0.8 \pm 0.2 ~\text{GeV}^2$~\cite{Belyaev:1982cd}\\
        $m_{\rho_3}$ & $1.76 \pm 0.01~\text{GeV}$~\cite{Aliev:2023tqy} \\
        $m_{\omega_3}$ & $1.76 \pm 0.01~\text{GeV}$~\cite{Aliev:2023tqy} \\
        $m_{K_3}$ & $1.82 \pm 0.02~\text{GeV}$~\cite{Aliev:2023tqy} \\
        $m_{\phi_3}$ & $1.86 \pm 0.01~\text{GeV}$~\cite{Aliev:2023tqy} \\
        $f_{\rho_3}$ & $(1.17 \pm 0.01) \times 10^{-2}$~\cite{Aliev:2023tqy}\\
        $f_{K_3}$ & $(1.26 \pm 0.01) \times 10^{-2}$~\cite{Aliev:2023tqy} \\
        $f_{\phi_3}$ & $(1.36 \pm 0.01) \times 10^{-2}$~\cite{Aliev:2023tqy} \\
        $\chi(1~\text{GeV})$ & $-(3.15 \pm 0.30)~\text{GeV}^{-2}$~\cite{Rohrwild:2007yt} \\
        $f_{3\gamma}$ & $-0.0039~\text{GeV}^2$~\cite{Ball:2002ps} \\
        \bottomrule
    \end{tabular}
    \label{tab:inputs}
\end{table}
%

The sum rules for the multipole moments of the $J^P = 3^-$ tensor mesons 
obtained in the previous section contain two more auxiliary parameters in
addition to the other input parameters, namely the Borel mass parameter $M^2$
and the continuum threshold $s_0$. The continuum threshold $s_0$ is determined
by requiring that the mass sum rules reproduce the mass  within 10\% accuracy.
Analysis of the mass sum rules performed in \cite{Aliev:2023tqy} shows that
this condition is satisfied  when $s_0$ varies in the region presented
in Table~\ref{tab:2}. On the other hand, the working region of the Borel mass
parameter is obtained from the following two conditions:
\begin{itemize}
\item[a)] The upper bound of $M^2$ is obtained by requiring that the pole
condition should dominate over contributions from higher states
and continuum.
\item[b)] The minimum value of $M^2$ is determined by demanding that the OPE
should be convergent. Borel masses that fulfill these requirements are
listed in Table \ref{tab:2}.
\end{itemize}

%
\begin{table}[hbt]
\begin{adjustbox}{center}
\renewcommand{\arraystretch}{1.2}
\setlength{\tabcolsep}{6pt}
  \begin{tabular}{lcc}
    \toprule
            &    $M^2\,(\rm{GeV}^2)$  & $s_0\,(\rm{GeV^2})$  \\
\midrule
$\rho_3$    &   $3.0 \div 5.0 $  & $5.25 \pm 0.25$  \\
$K_3$       &   $3.0 \div 5.0 $  & $5.75 \pm 0.25$  \\
    \bottomrule
  \end{tabular}
\end{adjustbox}
\caption{Working regions for the continuum threshold $s_0$
 Borel mass $M^2$. }
\label{tab:2}
\end{table}
%

\begin{figure}[h]
  \centering
\includegraphics[width=0.77\textwidth]{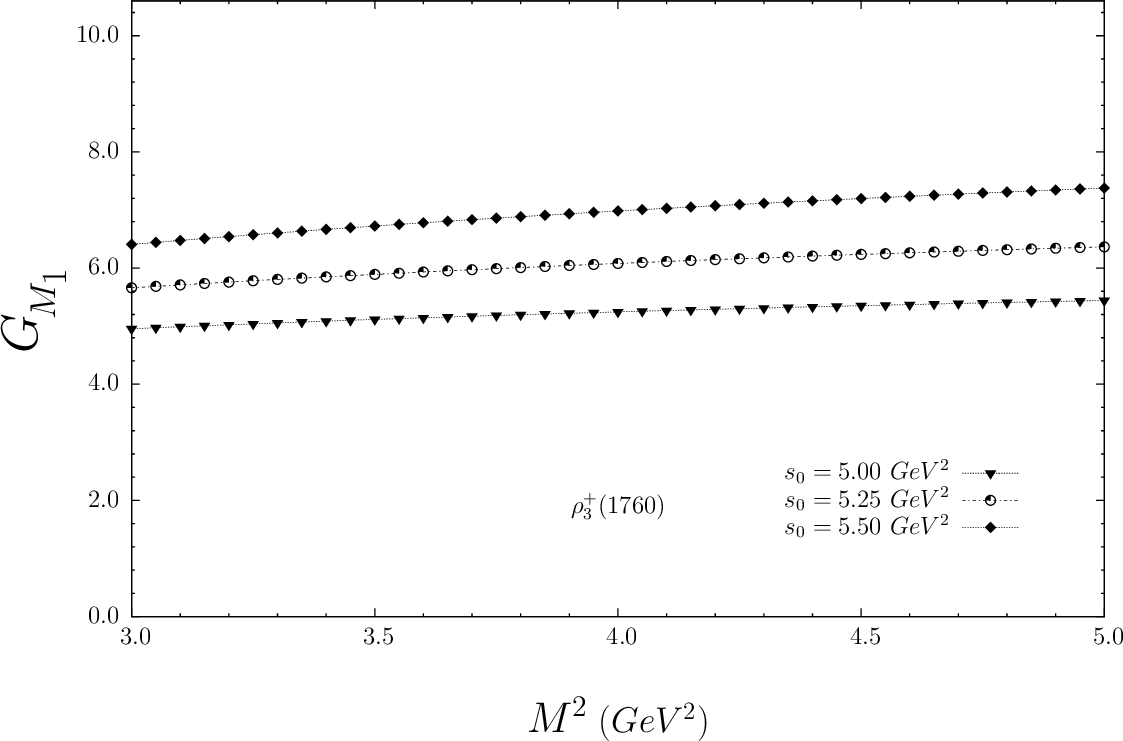}
\caption{
The dependence of the magnetic dipole moment $G_{M_1}$ of the $\rho_3^+$ tensor meson on $M^2$ at three fixed values of $s_0$.}
\label{fig:fig1}
\end{figure}

\begin{figure}[h]
  \centering
\includegraphics[width=0.77\textwidth]{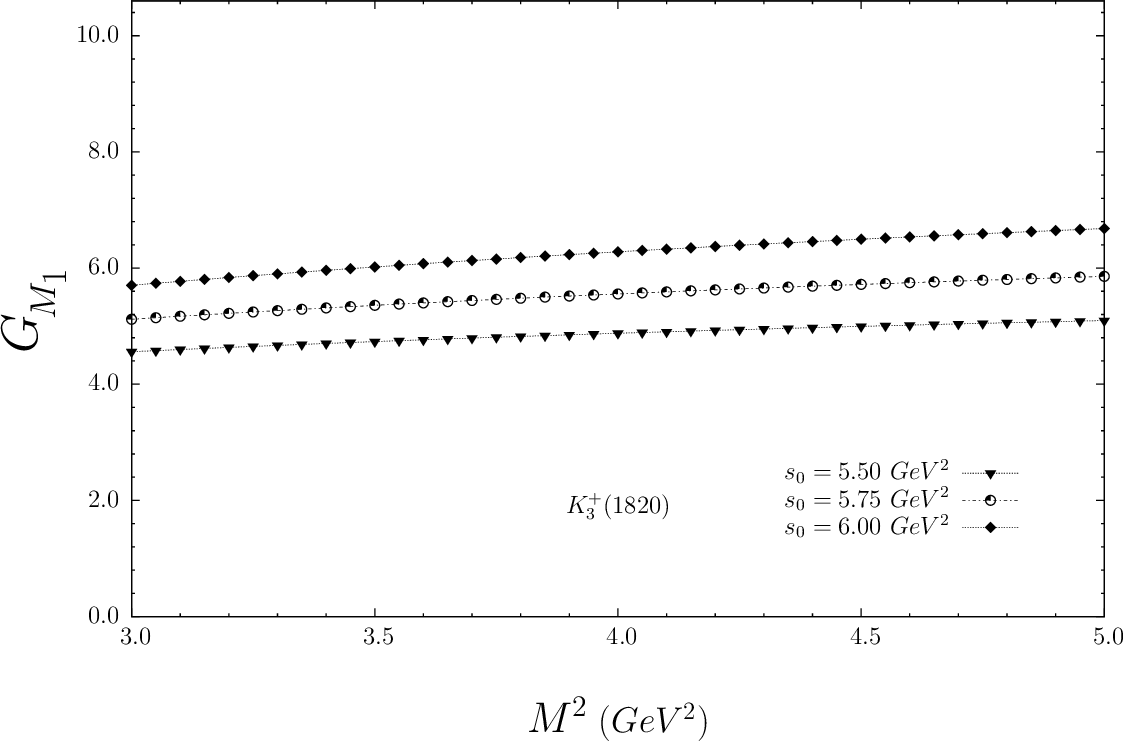} 
\caption{
The same as Fig. 1, but for the $K_3^+$ tensor meson.}
\label{fig:fig2}
\end{figure}

\begin{figure}[h]
  \centering
\includegraphics[width=0.77\textwidth]{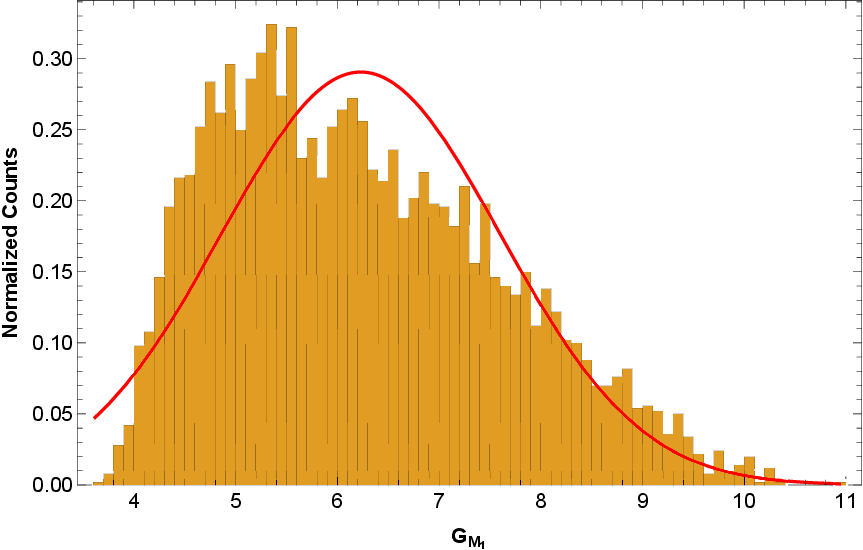} 
\caption{
The values of $G_{M_1}$ is calculated with random parameters within the uncertainty ranges of input paramters for 5000 points. The results have been fitted with Gaussian distribution. }
\label{fig:fig3}
\end{figure}

Having established the working regions of $M^2$ and $s_0$, we can now calculate multipole moments of the $J^P = 3^-$ tensor mesons using the
corresponding sum rules. As an example, in Figs. (\ref{fig:fig1}) and (\ref{fig:fig2}) we present
the dependence of $G_{M_1}$ on $M^2$ at several fixed values of $s_0$ for
$\rho_3^+$ and $K_3^+$ mesons, respectively and we observe good stability with respect to the variation of $M^2$ in its working domain

The uncertainties in the multipole moment values arise from a combination of factors, including uncertainties in the input parameters of photon Distribution Amplitudes (DAs), variations in the values of $s_0$ and $M^2$, as well as the errors in the values of the remaining input parameters. To study sensitivity of multipole moments to the uncertainties of all input parameters, we used the Monte-Carlo procedure which allows simultaneously take into account all uncertainties. (About this method see~\cite{Leinweber:1995fn}).  In Fig.~\ref{fig:fig3}, as an example, we present the histogram for $G_{M1}$ for $\rho_3^+$ meson. A similar analysis has been conducted for all multipole moments and the obtained results are presented in Table~\ref{tab:3}.


\begin{table}[hbt]
\begin{adjustbox}{width=\columnwidth,center}
\renewcommand{\arraystretch}{1.9}
\setlength{\tabcolsep}{6pt}
\small
\centering
  \begin{tabular}{lccccccc}
    \toprule
            & $G_{E_0}$
            & $G_{M_1}$
            & $G_{E_2}$ 
            & $G_{M_3}$
            & $G_{E_4}$
            & $G_{M_5}$
            & $G_{E_6}$ \\
\midrule
$\rho_3^+$     & $0.72 \pm 0.16$  & $ 6.28 \pm 1.40 $  & $-12.87  \pm 2.78$  &
$-18.07 \pm 3.92 $ & $18.29 \pm 4.04$  & $6.39 \pm 1.54$  & $  -0.43 \pm 0.73$    \\
$K_3^+$        & $0.65 \pm 0.08$ & $5.59 \pm 0.72$  & $-11.40  \pm 1.41$  &
$ -16.41 \pm 1.97 $ & $16.62 \pm 2.20$  & $6.39 \pm 1.54 $  & $  -2.62 \pm 0.58$     \\
\midrule
\cite{Lorce:2009br} & 1  & 6  & -15  & -20  & 15  & 6  & 1   \\
    \bottomrule
  \end{tabular}
\end{adjustbox}
\caption{The multipole moments in units of $(e/m_3^{\ell})$ and 
$(e/2 m_3^{\ell})$ for $G_{E_\ell}$ and $G_{M_\ell}$,
respectively.} 
\label{tab:3}
\end{table}
It is important to note that these multipole moments were previously estimated in \cite{Lorce:2009br} using the light-cone helicity amplitude framework. For completeness, Table \ref{tab:3} presents the results for the multipole moments of the charged $\rho_3^+$ and $K_3^+$ mesons calculated in \cite{Lorce:2009br}, where the effects of $SU(3)$ symmetry violation were neglected. Upon comparing our results with those in \cite{Lorce:2009br}, we find that the predictions for the multipole moments in both approaches are consistent within the uncertainties. The discrepancies observed can likely be attributed to the $SU(3)$ symmetry violation effects omitted in \cite{Lorce:2009br}, as well as uncertainties in the input parameters. Additionally, we estimated the multipole moments for the neutral $K_3^0$ meson and found that the maximum $SU(3)$ symmetry violation is around 15\%.

In conclusion, we have calculated the multipole moments of $J^P=3^-$ tensor mesons using the light-cone sum rules method. We compared our results with those obtained in \cite{Lorce:2009br} and observed that the predictions from both approaches are in close agreement. The differences can be traced back to uncertainties in the input parameters and the effects of $SU(3)$ symmetry violation. Our results provide valuable predictions for the multipole moments of these mesons, which can be tested in future experiments.



\bibliographystyle{utcaps_mod}
\bibliography{all.bib}


\newpage


%
\appendix
\section{Invariant functions}
\label{appendix}
In this appendix, we present the expressions of the invariant functions in
various structures.
\subsection{Coefficient of the structure 
$({\cal P}\!\cdot\!\varepsilon) g_{\alpha\rho} g_{\beta\nu} g_{\mu\sigma}$ }
\bea
\Pi_{1} \es
- {1\over 20160 \pi^2} (e_{q_1} - e_{q_2}) (2 {\cal I}_3 - 7 {\cal I}_2 m_{q_1}
m_{q_2})~. 
\eea
\subsection{Coefficient of the structure
$({\cal P}\!\cdot\!\varepsilon) q_{\rho} q_{\sigma} g_{\alpha\nu} g_{\beta\mu}$ }
\bea
%
\Pi_{2} \es
- {1\over 120960 \pi^2}
 \Big\{e_{q_1} \Big[12 {\cal I}_2 + 7 m_{q_1} \Big({\cal I}_1 m_{q_2} + 
     30 m_0^2 \pi^2 \langle \bar{q}_2 q_2 \rangle\Big)\Big] \nnb \\ 
\ek e_{q_2} \Big[12 {\cal I}_2 + 7 m_{q_2} \Big({\cal I}_1 m_{q_1} +
       30 m_0^2 \pi^2 \langle \bar{q}_1 q_1 \rangle\Big)\Big]\Big\}~. 
\eea
\subsection{Coefficient of the structure
$({\cal P}\!\cdot\!\varepsilon) q_{\beta} q_{\nu}q_{\rho} q_{\sigma} g_{\alpha\nu}$ } 
%
\bea
\Pi_{3} \es
- {1\over 144} \Big[ f_{3\gamma} (e_{q_1} - e_{q_2}) {\cal I}_0 
\widetilde{j}_1^0(\psi^v) \Big] \nnb \\
\ar {1\over 864}
  e_{q_1} m_{q_2} \langle \bar{q}_1 q_1 \rangle \Big\{ 3 \widetilde{j}_2^1(h_\gamma) +
4\Big[ 2 \widetilde{j}_2^2(h_\gamma) + \widetilde{j}_2^3(h_\gamma)\Big]\Big\}\nnb \\
\ek {1\over 432}  e_{q_1} f_{3\gamma} {\cal I}_0 \Big\{11 \widetilde{j}_2^1(\psi^v) +
     4 \Big[3 \widetilde{j}_2^2(\psi^v) +
\widetilde{j}_2^3(\psi^v)\Big]\Big\} \nnb \\
\ar {1\over 864}
 e_{q_2} m_{q_1} \langle \bar{q}_2 q_2 \rangle \Big\{ 3 \widetilde{j}_3^1(h_\gamma) -
4\Big[ 2 \widetilde{j}_3^2(h_\gamma) - \widetilde{j}_3^3(h_\gamma)\Big]\Big\}\nnb \\
\ek {1\over 432}  e_{q_2} f_{3\gamma} {\cal I}_0 \Big\{11 \widetilde{j}_3^1(\psi^v) -
     4 \Big[3 \widetilde{j}_3^2(\psi^v) -
\widetilde{j}_3^3(\psi^v)\Big]\Big\} -
{1\over 1152} e_{q_1} f_{3\gamma} {\cal I}_0 \psi^a(u_0) \nnb \\
\ar {1\over 27648 M^2} m_0^2 \Big[e_{q_1} \langle \bar{q}_2 q_2 \rangle
( 96 m_{q_1} -  103 m_{q_2}) + e_{q_2} \langle \bar{q}_1 q_1 \rangle
( 103 m_{q_1} - 96 m_{q_2}) \Big] \nnb \\
\ek {1\over 64512 \pi^2} e_{q_1}  {\cal I}_1 +
{1\over 138240 \pi^2} e_{q_1} m_{q_1}  \Big[23 m_{q_2} {\cal I}_0 - 180 \pi^2
\langle \bar{q}_2 q_2 \rangle\Big] \nnb \\ 
\ek {23\over 138240 \pi^2} e_{q_2} m_{q_1} m_{q_2} {\cal I}_0 \\ \nnb
\ar
{1\over 193536 \pi^2} e_{q_2} \Big\{3 {\cal I}_1 + 28 \pi^2 \Big[ 6 f_{3\gamma} {\cal I}_0 \psi^a(1-u_0)
 + 9 m_{q_2} \langle \bar{q}_1 q_1 \rangle\Big]\Big\}~. 
\eea
%
\subsection{Coefficient of the structure
$({\cal P}\!\cdot\!\varepsilon) q_{\alpha} q_{\beta}q_{\mu} q_{\nu} q_{\rho}
q_{\sigma}$ }
%
\bea
\Pi_{4} \es
{9\over 35840 \pi^2} e_{q_1} {\cal I}_0 -
{1\over 1536 M^4} e_{q_1} \Big\{ (35 m_0^2 - 18 M^2) m_{q_2} \langle \bar{q}_2 q_2 \rangle \nnb \\
\ar      4  f_{3\gamma} M^4 \Big[9 \widetilde{j}_1^0(\psi^v) + 8 \Big(6 \widetilde{j}_2^1(\psi^v) +
          11 \widetilde{j}_2^2(\psi^v) + 8 \widetilde{j}_2^3(\psi^v) + 2
\widetilde{j}_2^4(\psi^v)\Big)\Big]\Big\} \nnb \\
\ek {9\over 35840 \pi^2} e_{q_2} {\cal I}_0 +
{1\over 1536 M^4} e_{q_2}  \Big\{(35 m_0^2 - 18 M^2) m_{q_1} \langle \bar{q}_1 q_1 \rangle \nnb \\
 \ar    4  f_{3\gamma} M^4 \Big[9 \widetilde{j}_1^0(\psi^v) - 8 \Big(6 \widetilde{j}_3^1(\psi^v)
- 11 \widetilde{j}_3^2(\psi^v) + 8 \widetilde{j}_3^3(\psi^v) - 2
\widetilde{j}_3^4(\psi^v)\Big)\Big]\Big\}~. 
\eea
%
\subsection{Coefficient of the structure
$\varepsilon_{\sigma} q_{\rho} g_{\alpha\nu} g_{\beta\mu}$ }
%
\bea
\Pi_{5} \es
{1\over 60480 \pi^2}
(e_{q_1} - e_{q_2}) \Big(2 {\cal I}_3 - 7 {\cal I}_2 m_{q_1} m_{q_2}\Big)~. 
\eea
\subsection*{ Coefficient of the structure
$\varepsilon_{\sigma} q_{\rho} g_{\alpha\beta} g_{\mu\nu}$ }
\bea
\Pi_{6} \es
{1 \over 362880 \pi^2} \Bigg\{
e_{q_2} \Big[ 37 {\cal I}_3 + 35 m_{q_1} (-3 {\cal I}_2 m_{q_2} +
     4 m_0^2 \pi^2 {\cal I}_0 \langle \bar{q}_1 q_1 \rangle) \Big] \nnb \\
\ek e_{q_1} \Big[ 37 {\cal I}_3 - 35 m_{q_2} (3 {\cal I}_2 m_{q_1} -
     4 m_0^2 \pi^2 {\cal I}_0 \langle \bar{q}_2 q_2 \rangle) \Big] \Bigg\} ~.
\eea
\subsection{ Coefficient of the structure
$\varepsilon_{\rho} q_{\sigma} g_{\alpha\beta} g_{\mu\nu}$ }
\bea
\Pi_{7} \es
- {1\over 120960 \pi^2} \Bigg\{
e_{q_2} \Big[11 {\cal I}_3 - 35 m_{q_1} ({\cal I}_2 m_{q_2} +
     4 m_0^2 \pi^2 {\cal I}_0 \langle \bar{q}_1 q_1 \rangle) \Big] \nnb \\
\ek e_{q_1} \Big[ 11 {\cal I}_3 - 35 m_{q_2} ({\cal I}_2 m_{q_1} +
     4 m_0^2 \pi^2 {\cal I}_0 \langle \bar{q}_2 q_2 \rangle) \Big]~.
\eea
%
\subsection*{Coefficient of the structure
$\varepsilon_{\sigma} q_{\beta} q_{\nu}q_{\rho} g_{\alpha\mu}$ }
%
\bea
\Pi_{8} \es
{1\over 384} \Big\{{\cal I}_0 \Big[e_{q_1} m_{q_2} \langle \bar{q}_1 q_1 \rangle -
e_{q_2} m_{q_1} \langle \bar{q}_2 q_2 \rangle\Big]
   \widetilde{j}_1^0(h_\gamma)\Big\}
 - {1\over 288} \Big[5  f_{3\gamma} (e_{q_1} - e_{q_2}) {\cal I}_1
   \widetilde{j}_1^0(\psi^v)\Big] \nnb \\
\ar {1\over 1728} e_{q_1} \Big\{{\cal I}_0 m_{q_2} \langle \bar{q}_1 q_1 \rangle
\Big[21 \widetilde{j}_2^1(h_\gamma) +
     6 \widetilde{j}_2^2(h_\gamma) - 4 \widetilde{j}_2^3(h_\gamma)\Big] -
   8  f_{3\gamma} {\cal I}_1 \Big[7 \widetilde{j}_2^1(\psi^v) +
   3 \widetilde{j}_2^2(\psi^v) \Big]\Big\} \nnb \\
\ar {1\over 1728} e_{q_2} \Big\{{\cal I}_0 m_{q_1} \langle \bar{q}_2 q_2 \rangle 
\Big[21 \widetilde{j}_3^1(h_\gamma) -
     6 \widetilde{j}_3^2(h_\gamma) - 4 \widetilde{j}_3^3(h_\gamma) \Big] -
   8  f_{3\gamma} {\cal I}_1 \Big[ 7 \widetilde{j}_3^1(\psi^v) -
3 \widetilde{j}_3^2(\psi^v) \Big] \Big\} \nnb \\
\ek {13\over 322560 \pi^2} e_{q_1} {\cal I}_2 - 
{1\over 69120 \pi^2}  e_{q_1} m_{q_1} \Big[{\cal I}_1 m_{q_2} -
      150 m_0^2 \pi^2 \langle \bar{q}_2 q_2 \rangle\Big] \nnb \\
\ek {1\over 6912}  e_{q_1} \Big[18  f_{3\gamma} {\cal I}_1 \psi^v(u_0) 
- 12 \mathbb{A}(u_0) {\cal I}_0 m_{q_2} \langle \bar{q}_1 q_1 \rangle
+ 13 m_0^2 m_{q_2} \langle \bar{q}_2 q_2 \rangle \Big] \nnb \\
\ar {13\over 322560 \pi^2} e_{q_2} {\cal I}_2 +
{1\over 2304} e_{q_2} \Big[ 6  f_{3\gamma} {\cal I}_1 \psi^v(1-u_0) -
        5 m_0^2 m_{q_2} \langle \bar{q}_1 q_1 \rangle \Big] \nnb \\
\ar {1\over 69120 \pi^2 } e_{q_2} m_{q_1} \Big\{ {\cal I}_1 m_{q_2} +
        10 \pi^2 \Big[13 m_0^2 \langle \bar{q}_1 q_1 \rangle -
12 \mathbb{A}(1-u_0) {\cal I}_0 \langle \bar{q}_2 q_2 \rangle\Big] \Big\}~.
\eea
%
\subsection{Coefficient of the structure
$\varepsilon_{\rho} q_{\beta} q_{\nu}q_{\sigma} g_{\alpha\mu}$ }
%
\bea
\Pi_{9} \es
 - {1\over 1152} {\cal I}_0 \Big[e_{q_1} m_{q_2} \langle \bar{q}_1 q_1 \rangle -
e_{q_2} m_{q_1} \langle \bar{q}_2 q_2 \rangle\Big]
   \widetilde{j}_1^0(h_\gamma) - 
 {7\over 864}  f_{3\gamma} (e_{q_1} - e_{q_2}) {\cal I}_1
   \widetilde{j}_1^0(\psi^v) \nnb \\
\ar {1\over 1728} {\cal I}_0 e_{q_1} m_{q_2} \langle \bar{q}_1 q_1 \rangle
\Big[ 3 \widetilde{j}_2^1(h_\gamma) +
     14 \widetilde{j}_2^2(h_\gamma) + 4 \widetilde{j}_2^3(h_\gamma) \Big] -
   {1\over 216} e_{q_1} f_{3\gamma} {\cal I}_1 \Big[ 5 \widetilde{j}_2^1(\psi^v) +
   3 \widetilde{j}_2^2(\psi^v) \Big] \nnb \\
\ar {1\over 1728} {\cal I}_0 e_{q_2} m_{q_1} \langle \bar{q}_2 q_2 \rangle
\Big[ 3 \widetilde{j}_3^1(h_\gamma) - 14 \widetilde{j}_3^2(h_\gamma) + 
4 \widetilde{j}_3^3(h_\gamma) \Big] -
{1\over 216}e_{q_2}  f_{3\gamma} {\cal I}_1 
\Big[ 5 \widetilde{j}_3^1(\psi^v) - 3 \widetilde{j}_3^2(\psi^v) \Big] \nnb \\
\ar {1\over 322560 \pi^2} e_{q_1} {\cal I}_2 
+ {1\over 69120 \pi^2}  e_{q_1} m_{q_1} \Big[ 29 {\cal I}_1 m_{q_2} -
      30 m_0^2 \pi^2 \langle \bar{q}_2 q_2 \rangle \Big] \nnb \\
\ek {1\over 6912} e_{q_1} \Big[6  f_{3\gamma} {\cal I}_1 \psi^v(u_0)
+ 12 \mathbb{A}(u_0) {\cal I}_0 m_{q_2} \langle \bar{q}_1 q_1 \rangle +
7 m_0^2 m_{q_2} \langle \bar{q}_2 q_2 \rangle \Big] \nnb \\
\ek {1\over 322560 \pi^2} e_{q_2} {\cal I}_2 +
{1\over 2304} e_{q_2} \Big[2  f_{3\gamma} {\cal I}_1 \psi^v(1-u_0) +
      m_0^2 m_{q_2} \langle \bar{q}_1 q_1 \rangle\Big] \nnb \\
\ek {1\over 69120 \pi^2} e_{q_2} m_{q_1} \Big\{ 29 {\cal I}_1 m_{q_2} -
      10 \pi^2 \Big[ 7 m_0^2 \langle \bar{q}_1 q_1 \rangle +
12 \mathbb{A}(1-u_0) {\cal I}_0 \langle \bar{q}_2 q_2 \rangle \Big] \Big\}~. 
\eea
\subsection{ Coefficient of the structure
$\varepsilon_{\rho} q_{\alpha} q_{\beta} q_{\sigma} g_{\mu\nu}$ }
\bea
\Pi_{10} \es
{1\over 144}
{\cal I}_0 \Bigg\{e_{q_1} m_{q_2} \langle \bar{q}_1 q_1 \rangle \Big[\widetilde{j}_2^1(h_\gamma) +
    4 \widetilde{j}_2^2(h_\gamma) + 4 \widetilde{j}_2^3(h_\gamma) \Big] \nnb \\
\ar  e_{q_2} m_{q_1} \langle \bar{q}_2 q_2 \rangle \Big[\widetilde{j}_3^1(h_\gamma) - 4 j\widetilde{j}_3^2(h_\gamma) +
    4 \widetilde{j}_3^3(h_\gamma) \Big] \Bigg\} \nnb \\
\ar {1\over 1152} \Bigg\{
e_{q_2} \Big[ 9 m_0^2 (m_{q_1} + m_{q_2}) \langle \bar{q}_1 q_1 \rangle + \mathbb{A}(1-u_0) {\cal I}_0 m_{q_1}
    \langle \bar{q}_2 q_2 \rangle \Big] \nnb \\
\ek e_{q_1} \Big[\mathbb{A}(u_0) {\cal I}_0 m_{q_2} \langle \bar{q}_1 q_1 \rangle +
   9 m_0^2 (m_{q_1} + m_{q_2}) \langle \bar{q}_2 q_2 \rangle \Big] \Bigg\} \nnb \\
\ek {1\over 80640 \pi^2}
 ( e_{q_1} - e_{q_2}) (26 {\cal I}_2 - 105 {\cal I}_1 m_{q_1} m_{q_2} )~.
\eea
%
\subsection{Coefficient of the structure
$\varepsilon_{\sigma} q_{\alpha} q_{\beta}q_{\mu} q_{\nu} q_{\rho}$ }
%
\bea
\Pi_{11} \es
{1\over 256} \Big\{ \Big[e_{q_1} m_{q_2} \langle \bar{q}_1 q_1 \rangle - e_{q_2} m_{q_1} 
\langle \bar{q}_2 q_2 \rangle\Big] \widetilde{j}_1^0(h_\gamma) -
  17  f_{3\gamma} (e_{q_1} - e_{q_2}) {\cal I}_0 
\widetilde{j}_1^0(\psi^v) \Big\} \nnb \\
\ar {1\over 2304} e_{q_1} \Big\{ m_{q_2} \langle \bar{q}_1 q_1 \rangle
\Big[ 39 \widetilde{j}_2^1(h_\gamma) +
     8 \Big( 5 \widetilde{j}_2^2(h_\gamma) - 3 \widetilde{j}_2^3(h_\gamma) - 6 \widetilde{j}_2^4(h_\gamma) -
       2 \widetilde{j}_2^5(h_\gamma) \Big) \Big] \nnb \\
\ek 8  f_{3\gamma} {\cal I}_0 \Big[66 \widetilde{j}_2^1(\psi^v) +
     67 \widetilde{j}_2^2(\psi^v) + 24 \widetilde{j}_2^3(\psi^v) + 2
\widetilde{j}_2^4(\psi^v) \Big] \Big\} \nnb \\
\ar {1\over 2304} e_{q_2} \Big\{m_{q_1} \langle \bar{q}_2 q_2 \rangle \Big[39 \widetilde{j}_3^1(h_\gamma) -
     8 \Big( 5 \widetilde{j}_3^2(h_\gamma) + 3 \widetilde{j}_3^3(h_\gamma) - 6 \widetilde{j}_3^4(h_\gamma) +
       2 \widetilde{j}_3^5(h_\gamma) \Big) \Big] \nnb \\ 
\ek 8  f_{3\gamma} {\cal I}_0 \Big[ 66 \widetilde{j}_3^1(\psi^v) -
     67 \widetilde{j}_3^2(\psi^v) + 24 \widetilde{j}_3^3(\psi^v) -
2 \widetilde{j}_3^4(\psi^v) \Big] \Big\} \nnb \\
\ek {289\over 215040 \pi^2} e_{q_1} {\cal I}_1
- {5\over 576 \pi^2}  e_{q_1} m_{q_1} m_{q_2} {\cal I}_0 \nnb \\
\ek {1\over 768} e_{q_1} \Big\{ f_{3\gamma} {\cal I}_0 \Big[ 8 \psi^a(u_0) + 3 \psi^v(u_0) \Big]
- m_{q_2} \Big[ 8 \mathbb{A}(u_0) - 3 \chi {\cal I}_0 \varphi_\gamma(u_0)
\Big] \langle \bar{q}_1 q_1 \rangle \Big\} \nnb \\
\ar {1\over 9216 M^2} e_{q_1} \Big[ (87 m_0^2 - 36 M^2) m_{q_1} -
2 (67 m_0^2 - 27 M^2) m_{q_2} \Big] \langle \bar{q}_2 q_2 \rangle \nnb \\
\ar  {289\over 215040 \pi^2} e_{q_2} {\cal I}_1 
+ {5\over 576 \pi^2} e_{q_2} m_{q_1} m_{q_2} {\cal I}_0 \nnb \\
\ar {1\over 3072 M^2} e_{q_2} \Big\{4  f_{3\gamma} M^2 {\cal I}_0 
\Big[8 \psi^a(1-u_0) + 3 \psi^v(1-u_0)\Big] -
      (29 m_0^2 - 12 M^2) m_{q_2} \langle \bar{q}_1 q_1 \rangle \Big\} \nnb \\
\ar {1\over 4608 M^2} e_{q_2} m_{q_1} \Big\{ (67 m_0^2 - 27 M^2) \langle \bar{q}_1 q_1 \rangle -
        6 M^2 \Big[8 \mathbb{A}(1-u_0) - 3 \chi {\cal I}_0
\varphi_\gamma(1-u_0)\Big] \langle \bar{q}_2 q_2 \rangle \Big\} \nnb \\
\ar {3\over 1024 \pi^2} (e_{q_1} - e_{q_2}) m_{q_1} m_{q_2} {\cal I}_{\ell n}~. 
\eea
The functions $\widetilde{j}_\ell(f(u))$, where
$(\ell=1,2,3)$; and
${\cal I}_n$, ${\cal I}_{\ell n}$ entering into the correlation functions $\Pi_n$ 
are defined as:
\bea
\label{nolabel}
\widetilde{j}_1(f(u)) \es \int_{u_0}^1 du f(u)~, \nnb \\
\widetilde{j}_2(f(u)) \es \int_{u_0}^1 du (u-u_0) f(u)~, \nnb \\
\widetilde{j}_3(f(u)) \es \int_{u_0}^1 du [u-(1-u_0)] f(u)~, \nnb \\
{\cal I}_n \es \int_{(m_{q_1} + m_{q_2})^2}^{\infty} ds\, s^n e^{-s/M^2}~,\nnb \\
{\cal I}_{\ell n} \es \int_{(m_{q_1} + m_{q_2})^2}^{s_0} ds\,  
 e^{-s/M^2}\left[ \gamma_E + \ln\left({s\over \Lambda^2}\right) \right]~.\nnb
\eea
Here, $f(u)$ is the generic notation of photon distribution amplitudes. Explicit expressions of the DA's are already obtained in~\cite{Ball:2002ps}. Hence, for brevity we do not present their explicit expressions here. In expression $\mathcal{I}_{l n}$, $\gamma_E$ is the Euler constant and $\Lambda$ is the QCD scale parameter. For numerical calculations, we use $\Lambda = 0.5~\text{GeV}$. 

\newpage

\end{document}